**Title:** Comments on "On the clinical meaningfulness of a treatment's effect on a time-to-event variable"

**Authors:** Christos Argyropoulos MD, PhD

**Affiliation:** Department of Internal Medicine, Division of Nephrology, University of New Mexico

**Corresponding Author:**

>Department of Internal Medicine,
>
>Division of Nephrology,
>
>University of New Mexico,
>
>1 University of New Mexico
>
>MSC 04-2785
>
>Albuquerque, New Mexico 87131,
>
>Phone: (505) 272-0600
>
>Email: cargyropoulos@salud.unm.edu

**Abstract**

Some years ago, Snapinn and Jiang[1] considered the interpretation and pitfalls of absolute versus relative treatment effect measures in analyses of time-to-event outcomes. Through specific examples and analytical considerations based solely on the exponential and the Weibull distributions they reach two conclusions: 1) that the commonly used criteria for clinical effectiveness, the ARR (Absolute Risk Reduction) and the median (survival time) difference (MD) directly contradict each other and 2) cost-effectiveness depends only the hazard ratio(HR) and the shape parameter (in the Weibull case) but not the overall baseline risk of the population. Though provocative, the first conclusion does not apply to either the two special cases considered or even more generally, while the second conclusion is strictly correct only for the exponential case. Therefore, the implication inferred by the authors i.e. all measures of absolute treatment effect are of little value compared with the relative measure of the hazard ratio, is not of general validity and hence both absolute and relative measures should continue to be used when appraising clinical evidence.

**Concordant views of clinical meaningfulness for the exponential distribution**

The major argument supporting a contradiction between the ARR and the MD is based on Table II [1] which shows that for the exponential distribution the ARR at any fixed point in time and a given HR increases, while the MD decreases for increasing hazard rate. This argument is problematic on two grounds: first and foremost, in a comparative effectiveness setup the relevant quantity should be the behavior of an alternative treatment effect measure as a function of the *hazard ratio* rather than the hazard rate as proposed in the manuscript. If two or more treatment effect measures are internally consistent, then whenever one increases (hazard ratio) so should the other (MD). Secondly, even if one adopts the authors' viewpoint, the assertion that the ARR will behave in a monotonic fashion does not hold true for either the exponential or more general proportional hazards distributions. To illustrate these points I will use the authors' notation, in which $\lambda$ stands for the HR (assumed to be less than 1 as in [1]) and $\gamma$ for the control event rate, to write the ARR and MD as:

$$ARR = e^{-\lambda \gamma t} - e^{-\gamma t}$$

$$MD = \frac{\ln 2}{\gamma \lambda} - \frac{\ln 2}{\gamma}$$

Note that the partial derivative of ARR with respects to $\lambda$ is always negative, so that both the ARR and the MD[1] are strictly decreasing functions of the HR as can be seen in the numerical examples presented in Table II[1]. This holds true irrespective of the value of the hazard rate and thus no inconsistency among the two measures of treatment efficacy exists. Even if one believes that treatment effect measures should be compared vis-à-vis their dependency on the baseline hazard rate, a straightforward partial differentiation of the equation of the ARR reveals that this quantity is not a strictly increasing function of the hazard rate as suggested by the numerical examples in Table II [1]:

$$\frac{d}{d\gamma} ARR = t\left(e^{-\gamma t} - \lambda e^{-\lambda \gamma t}\right)$$

In particular, ARR will have a local maximum when $\gamma t = \frac{\ln \lambda}{\lambda - 1}$ and thus will be a strictly increasing function when $\gamma < \gamma_{crit} = \frac{\ln \lambda}{t(\lambda - 1)}$ but strictly decreasing otherwise. By judicial selection of the time at which to compute the ARR, the ARR can be made a strictly decreasing function for any value of the hazard rate and thus brought into agreement with the behavior of the MD. Specifically, substituting for $t$ any value larger than $t_* = \frac{\ln \lambda}{\gamma(\lambda - 1)}$, the point at which the ARR obtains its maximum value as a function of time (Table I[1]) will lead to a strictly decreasing ARR. Therefore, the apparent inconsistency between the ARR and the MD is only evident if one were to consider a limited range of baseline hazard rates and observation times [1].

**The general proportional hazards case**

The arguments of the previous paragraph also hold for a general proportional hazards distribution with a continuous hazard rate function. In the general case, the constant baseline hazard rate $\gamma$ becomes a function of time $\gamma(t)$; for any specific time $t_S$ at which to calculate the ARR, the first mean value theorem for integration allows us to write the survival function for the control and experimental arms as:

$$\exp(-\int_0^{t_S} \gamma(t)dt) = \exp(-\overline{\gamma_S} t_S) \text{ and } \exp(-\lambda \int_0^{t_S} \gamma(t)dt) = \exp(-\lambda \overline{\gamma_S} t_S)$$

where $\overline{\gamma_S}$ is the average value of the hazard rate function in the interval $[0, t_S]$. Since the average hazard rate is positive, the partial derivative of the ARR in the general proportional hazards case is a decreasing function of the HR irrespective of the average rate. Furthermore, the ARR will exhibit a local maximum $\overline{\gamma_S} t_S = \frac{\ln \lambda}{\lambda - 1}$ for a fixed $t_S$ recapitulating the behavior of the exponential case.

Similar to the exponential distribution, the median difference is a decreasing function of both the hazard ratio and the baseline average rate for the general proportional hazard model. To see why note that the median survival time for the control arm ($t_{0.5}^C$) is the solution of the non-linear equation: $\int_0^{t_{0.5}^C} \gamma(t) dt = \overline{\gamma_{0.5}^C} t_{0.5}^C = \ln 2$, where $\overline{\gamma_{0.5}^C}$ is the associated average control hazard rate. For a protective ($\lambda < 1$) intervention the median survival time ($t_{0.5}^E$) will always exceed $t_{0.5}^C$ so that one may write: $\int_0^{t_{0.5}^E} \gamma(t) dt = \int_0^{t_{0.5}^C} \gamma(t) dt + \int_{t_{0.5}^C}^{t_{0.5}^E} \gamma(t) dt = \overline{\gamma_{0.5}^C} t_{0.5}^C + \overline{\Delta \gamma_{0.5}^C} (t_{0.5}^S - t_{0.5}^C)$. This expression follows by applying the mean value theorem twice and noting that all quantities appearing in the right-hand side are strictly positive. At the median survival time, one may write the following relation for the experimental arm:

$\lambda \int_0^{t_S} \gamma(t) dt = \ln 2 = \overline{\gamma_{0.5}^C} t_{0.5}^C \Rightarrow \lambda \overline{\gamma_{0.5}^C} t_{0.5}^C + \lambda \overline{\Delta \gamma_{0.5}^C} (t_{0.5}^S - t_{0.5}^C) = \overline{\gamma_{0.5}^C} t_{0.5}^C$. Hence the median difference $t_{0.5}^S - t_{0.5}^C$ is equal to $\frac{(1-\lambda) \overline{\gamma_{0.5}^C} t_{0.5}^C}{\lambda \overline{\Delta \gamma_{0.5}^C}} = \frac{(1-\lambda) \ln 2}{\lambda \overline{\Delta \gamma_{0.5}^C}}$ which is a strictly decreasing function of the hazard ratio and the average hazard rate, $\overline{\Delta \gamma_{0.5}^C}$, in the interval $[t_C, t_E]$.

*To summarize, a rather straightforward analytical investigation does not support the authors' claim that the absolute risk reduction and the median survival time difference contradict each other for either the special case of the exponential distribution or the more general proportional hazards model.*

**Number Needed to Treat**

Having established the lack of inconsistency between ARR and MD, I feel it is of some value to revisit the NNT (number needed to treat) example in two clinical trials: A (a cardiovascular trial with a relatively low background event rate) and B (an oncology trials with a high hazard rate) the authors considered. In both cases, the same value of the HR of the experimental intervention is assumed and the authors conclude that the NNTs calculated at year one (27 in trial A, 7 in trial B) lead to discordant conclusions about the "clinical meaningfulness" of the two interventions when contrasted against the MDs (5 years in trial A but 9 months in B). Yet this apparent discrepancy is entirely dependent on the choice of the time at which to compute the NNT. This is shown in Figure 1 which considers the entire range of observation times that the authors consider in the corresponding Kaplan Meier curves (0-12 years).Over this range of observation times, the NNT of trial A may be higher, equal or lower than the NTT of trial B and this relation is heavily influenced by the background rate of the control treatment in both trials. In particular the NNT in trial A after the 6$^{th}$ year is much smaller that the corresponding NNT for trial B, reflecting the "clinical meaningfulness" inferences that Snapinn and Jiang would presumably want us to make (i.e. the experimental treatment in Trial A is more "useful" than the one in B). This figure reveals the major problems with the NNT (and the ARR), namely their time and background rate dependency. One may attempt to address these dependencies by computing the NNT at a time that is normalized to the background rate, i.e., the point of maximum separation between the survival curves. $t_*$ that Snapinn and Jiang considered. In this case, the NNT becomes independent of the background rate and a direct, non-linear function of the HR. Consequently, the NNT does not convey any additional, let alone contradictory, information over the HR about the relative effectiveness of the experimental treatments *within* each trial.

**Cost effectiveness and the general proportional hazards case**

The computation of the incremental cost-effectiveness ratio (ICER) requires the calculation of the incremental cost to the incremental benefit. If the (unit) costs of the control and the experimental treatment are $c_0$ and $c_E$ the value of the ICER over the lifetime is equal to:

$$ICER = \frac{c_E MST_E - c_0 MST_0}{MST_E - MST_0} = \frac{c_E - c_0(MST_0/MST_E)}{1-(MST_0/MST_E)} \approx \frac{c_E}{1-(MST_0/MST_E)}, c_0 << c_E$$

The ICER is a function of the ratio of the mean survival time (MST) under each treatment and the MST can be computed by the Area Under the Curve (AUC) of the survival function. The ICER will not depend on the baseline hazard rate if and only if the MST ratio does not depend on the hazard rate. Even though this condition holds for the exponential distribution, it is only partly satisfied by the Weibull distribution, and it does not obtain for more general proportional hazards models as shown below.

The hazard of the Weibull distribution with shape parameter $k$ and scale parameter $s$ can be alternatively expressed as $skt^{k-1}$ or $s^k kt^{k-1}$ or $st^{k-1}$ or $s^{-k}kt^{k-1}$ depending on the parameterization. Irrespective of the parameterization adopted, the ratio of the ICER will not be free of the baseline hazard rate since the ratio of the MSTs (HR$^{1/k}$) is a function of the shape parameter. The dependency of Weibull's ICER on this factor can be given a rather intuitive interpretation: a given force (the value of the HR, known as *force of mortality* in actuarial science) will displace an object (lifetime of an individual) moving in the opposite direction

(disease process) a lot further (prolongation of survival) and at a lower energy cost (value of the ICER) if the object is decelerating ($k<1$, decreasing baseline hazard) instead of accelerating ($k>1$, increasing baseline hazard).

With $H(t)$ standing for the cumulative baseline hazard and $\lambda$ for the HR, the ratio of MSTs is given by:

$$\frac{MST_0}{MST_E} = \frac{\int_0^\infty S_0(t)dt}{\int_0^\infty S_E(t)dt} = \frac{\int_0^\infty S_0(t)dt}{\int_0^\infty S_0(t)^\lambda dt} = \frac{\int_0^\infty \exp(-H(t))dt}{\int_0^\infty \exp(-\lambda H(t))dt}$$

Standard asymptotic analysis can be used to prove that the ICER of general proportional hazard distributions that are in some sense "close to the exponential" will be a function of the baseline hazard for general proportional hazard distributions. The notion of "exponential closeness" can be operationalized by requiring that the limit of the cumulative hazard function close to the origin ($t=0$) be approximated by a linear function: $H(t) \to H_0 + H_1 t, \ t \to 0$ with $H_0, H_1,\ldots,H_n$ signifying the value of the cumulative hazard its first and higher order derivatives at $t=0$

Since the cumulative hazard steadily increases towards positive infinity, the integrand in the definition of the MST decreases steadily away from the lower limit of integration. Consequently it admits a, power series representation (Chapter 5 of [2]) which for the control arm is given by:

$$\int_0^\infty \exp(-H(t))dt = H_1^{-1} \sum_{r=0}^{r=\infty} L_r$$

The terms in the expansion can be derived in at various ways. The most useful one for the purpose of this exposition is through the mixed algebraic/differential operator:

$$L_r = H_1 \left(\frac{d}{H'dt}\right)^r \frac{1}{H'}\bigg|_{t=0}$$

This is an asymptotic, rather than convergent, series so that the first few terms may be used to approximate each of the integrals appearing in the MST ratio. A similar expression applies to the experimental arm, by incorporating the HR into the cumulative hazard. Keeping only the first three terms, the MST ratio is complex function of the HR($\lambda$) and the value of the first three derivatives of the cumulative hazard function at the origin:

$$\lambda \frac{\left(1 - \frac{H_2}{H_1^2} + \frac{3H_2^2}{H_1^4} - \frac{H_3}{H_1^3}\right)}{\left(1 - \frac{H_2}{\lambda H_1^2} + \frac{3H_2^2}{\lambda^2 H_1^4} - \frac{H_3}{\lambda^2 H_1^3}\right)}$$

Inclusion of higher order terms will make this ratio a function of higher powers of the hazard ratio as well as higher order derivatives of the cumulative hazard function. Nevertheless, high accuracy is assured with inclusion of up to 8 terms if $H_1/\sqrt{H_2}$ is large (page 115 in [2]). Using the HR as a proxy for the ratio of the mean survival times is equivalent to retaining only the first term in the asymptotic expansion for the MST. Higher accuracy requires one to consider higher order terms which are functions of the baseline hazard ($H_1$), its rate of change ($H_2$) and its concavity at the origin($H_3$). This analysis, while confirming the assertion by Snapinn and Jiang that neither the ARR nor the difference in median survival cannot be intuitively translated to differences in cost-effectiveness, it illustrates even in the "close to exponential" general proportional hazards case, the HR is only a very crude and numerically inaccurate estimate to the ratio of survival times. Hence in health economic evaluations, should not rely on single term exponential approximation that the HR provides, but combine the latter with estimates of the baseline hazard.

**Figures**

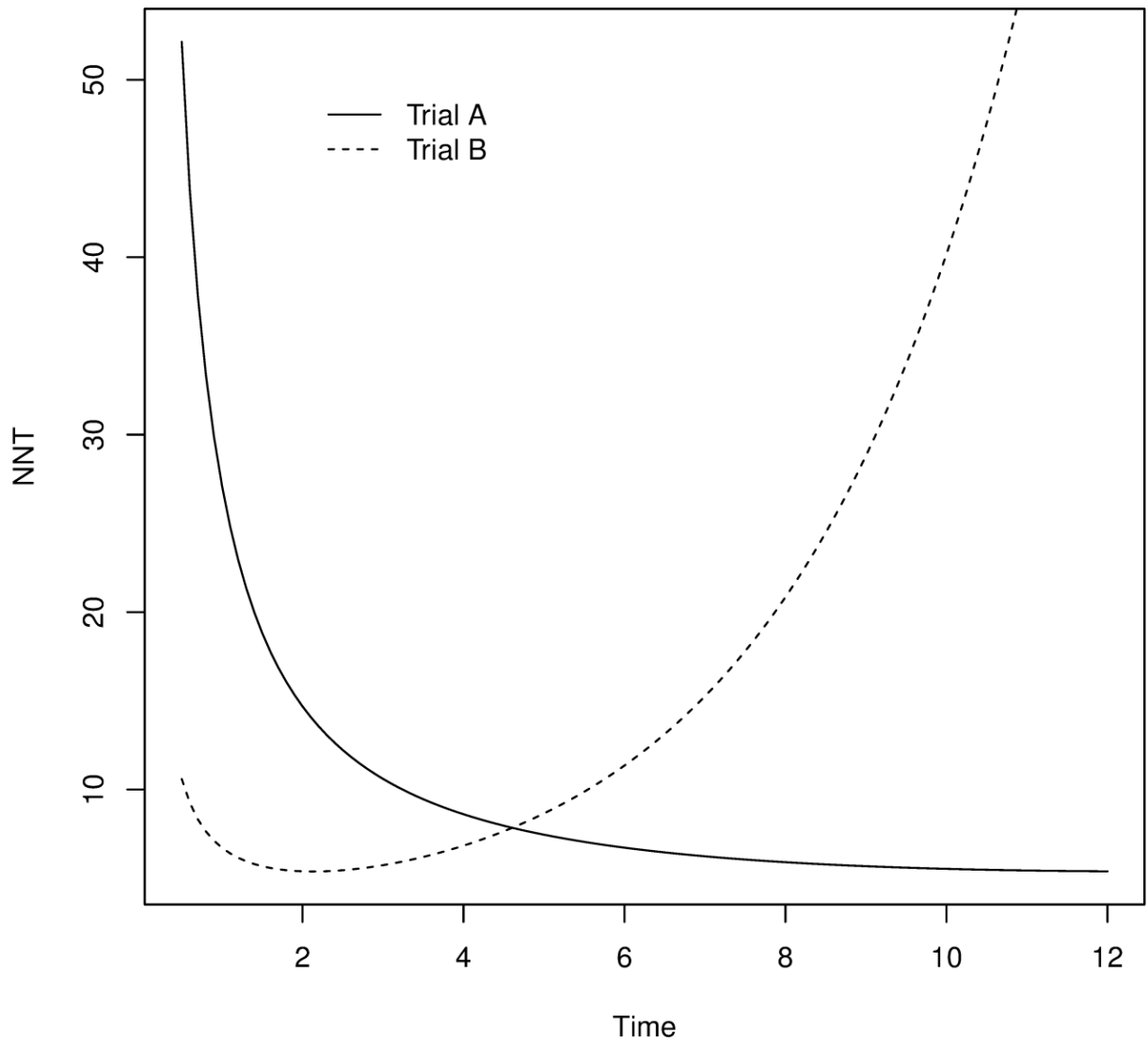

**Figure 1**. Number Needed to Treat (NNT) for a cardiology (Trial A) and an Oncology Trial (Trial B). The baseline hazard rate is 10% per year (Trial A) and 60% per year (trial B). The hazard ratio for the experimental intervention is 0.6 for both trials

**Data availability:**

Data sharing is not applicable to this article as no new data were created or analyzed in this study.